\documentstyle[amssymb,12pt]{article}

\begin{document}

\title{ RECURTION OPERATOR AND RATIONAL LAX REPRESENTATION}
\author{ K. Zheltukhin\\
 Department of Mathematics, Faculty of Sciences\\
 Bilkent University, 06533 Ankara, Turkey\\ 
Phone: 90 312 2901938 (office)\\
Fax~~: 90 312 2664579\\
e-mail: zhelt@fen.bilkent.edu.tr}

\date{}

\begin{titlepage}

\maketitle

\begin{abstract}
We  consider equations arising from  rational Lax
representations. A general method to construct
recursion operators for such equations is given.
Several examples are given, including a degenerate bi-Hamiltonian 
system with a recursion operator.

\medskip

{\it
\noindent
PACS:  0230.Ik; 0230.Sr \\
Keywords:  integrable system, recursion operator.
}

\end{abstract}

\end{titlepage}

\section*{I. Introduction}

Recently a new method of constructing a recursion operator
from Lax representation was introduced in \cite{GKS}. 
This construction depends on  Lax representation of a given system
of PDEs.  Let

\begin{equation}
L_t=[A,L]
\label{lax}
\end{equation}

\noindent
be  Lax representation of  an integrable nonlinear system of PDEs.
Then a hierarchy of symmetries can be given by

\begin{equation}
L_{t_n}=[A_n,L],\qquad n=0,1,2 \dots ,
\end{equation}

\noindent
where $t_0=t$, $A_0=A$ and $A_n,\quad n=0,1,2\dots,\quad$ are Gel'fand-Dikkii 
 operators given in terms of~$L$.
The recursion relation between symmetries can be written as
\begin{equation}
L_{t_{n+1}}=LL_{t_n} + [R_n,L],\qquad n=0,1,2 \dots,
\label{SymmRel1} 
\end{equation} 
where $R_n$ is an operator such that $ord R_n=ord L$.\\
This symmetry relation allows to find $R_n$, hence $L_{t_{n+1}}$,
in terms of $L$ and $L_{t_n}$.

In \cite{GKS}, \cite{B} this method was applied to construct
recursion operators for Lax equations with 
different classes of scalar and shift operators, corresponding to
field and lattice systems respectively.
In \cite{GZh} the method was applied to Lax equations on  a Poissson algebra of
Laurent series 

\begin{equation}
\Lambda = \left \{\sum_{-\infty}^{+\infty} u_ip^i\; : u_i
 \mbox{ -- smooth} \quad \mbox{functions}  \right \} 
\end{equation}

with  the polynomial Lax function. 
Such equations  give systems of hydrodynamic type.
They were also discussed in \cite{Str}-- \cite{BurD}.
The Hamiltonian structure of the Lax equation on a Poisson algebra
was studed in \cite{L}.

Here we  consider the Lax equation on the Poisson algebra $\Lambda$ with a 
rational Lax function
\begin{equation}
L=\frac{\Delta_1}{\Delta_2}\; ,
\label{LF}
\end{equation}
where $\Delta_1$, $\Delta_2$ are polynomials of degree $N$ and $M$, 
respectively, and  $N>M$.
The Lax equation is

\begin{equation}
 \frac{\partial L}{\partial t_n}=\{(L)^{\frac{1}{N-M}+n}_{\ge 0},L\} \; ,
\label{LaxEqn}
\end{equation}

where the Poisson bracket is given by
$$
\{f,g\}=p\left ( \frac{\partial f}{\partial p}
\frac{\partial g}{\partial x} - \frac{\partial f}{\partial x}
\frac{\partial g}{\partial p} \right ).
$$

 First we study the symmetry relation (\ref{SymmRel1}) for the
rational Lax function. Then we give some examples.

In particular, we find a recursion operator $\cal R$  for equation  
(\ref{LaxEqn}) with the Lax function

\begin{equation}
L=p+S+\frac{P}{p+Q} \; ,
\label{LF1}
\end{equation}

which leeds to the system \cite{Str} 

\begin{equation}
\begin{array}{lll}
S_t & = & P_x, \\ 
P_t & = & PS_x - QP_x - PQ_x, \\
Q_t & = & QS_x - QQ_x . 
\end{array}
\label{Eqn0}
\end{equation} 

The recursion operator is given by

\begin{equation}
{\cal R}=\left ( \matrix{
S  & 1   & PQ^{-1}+P_xD_x^{-1}\cdot Q \cr
2P & S-Q & -2P +(PS_x -(PQ)_x )D_x^{-1}\cdot Q\cr
Q & 1   & PQ^{-1}+S-Q +(QS_x - QQ_x)D_x^{-1}\cdot Q \cr
}\right )
\label{Rec1}
\end{equation} 

In \cite{Str} bi-Hamiltonian representation of this 
equation was constructed with Hamiltonian operators   

\begin{equation}
{{\cal D}_1}=\left ( \matrix{
0 & P & Q \cr
P & -2PQ & -Q^2 \cr
Q & -Q^2 & 0 \cr
}\right )D_x
+
\left ( \matrix{
0 & P_x    & Q_x  \cr
0 & -(PQ)_x & -QQ_x \cr
0 & -QQ_x  & 0     \cr
}\right )
\label{HamOper2}
\end{equation} 
and
\begin{equation}
{{\cal D}_2}=\left ( \matrix{
2P & P(S-3Q) & Q(S-Q) \cr
P(S-3Q) & P(2P-2SQ+4Q^2) & Q(2P-SQ +Q^2) \cr
Q(S-Q) & Q(2P-SQ +Q^2) & 2Q^2 \cr
}\right )D_x  + 
\label{HamOper1}
\end{equation}

$$
\left ( \matrix{
P_x & SP_x-2(PQ)_x & SQ_x-QQ_x \cr
PS_x-(QP)_x & (-SPQ + P^2+2PQ^2)_x & Q_x(2P + Q^2-SQ)\cr
QS_x-QQ_x & Q(2P_x+2QQ_x-S_x-SQQ_x)  & 2QQ_x \cr
}\right )
$$ 
These Hamiltonian operators are degenerate, so, 
one can not use them to find a recursion operator.
But it turns out that they are related to the recursion operator ${\cal R}$.
One can easily check that the following equality holds 
$$       
{\cal R}{\cal D}_1={\cal D}_2.
$$
We observe that the degeneracy in the bi-Hamiltonian operators is due to
the following fact. 
Let $p'=p+F$ then the Lax function becomes 

\begin{equation} 
L=p'+G+\frac{P}{p'} \; .
\label{mLF}
\end{equation}

This means that we have two independent variabels $P$ and $G$, where \\ 
$G=S-F$. The equation corresponding to the 
 Lax function (\ref{mLF}) has been studied in \cite{GZh}.

To remove  degeneracy 
one can take the Lax function as

\begin{equation} 
L=p+S+\frac{P}{p}+\sum_{i=1}^{m} \frac{Q_i}{p+F_i} \; . 
\end{equation}

As an example we shall consider the equation (\ref{LaxEqn}) with 
the Lax function

\begin{equation}
L=p+S+\frac{P}{p}+\frac{Q}{p+F}\; .
\label{LF2}
\end{equation} 

\section* {II. Symmetry Relation for Rational Lax Representation.}

\noindent
Following \cite{GKS} we consider the hierarchy of symmetries
for the Lax equation (\ref{LaxEqn}) with the Lax function (\ref{LF})

\begin{equation}       
\frac{\partial L}{\partial t_n}=\{ (L^{\frac{1}{N-M}+n})_{\ge 0}, L\}. 
\end{equation} 

\vspace{0.3cm}
  
\noindent
{\bf Lemma 1. } {\it
 For any $n=0,1,2,\dots, $
\begin{equation}       
\frac{\partial L}{\partial t_n}=L\frac{\partial L}{\partial t_{n-1}} 
+ \{R_n,L\}.
\label{SymmRel}
\end{equation}

\noindent
Function $R_n$ has a form

\begin{equation}
R_n= A+ \frac{B}{\Delta_2}
\label{FormRec}
\end{equation}

\noindent
where $A$ is a polynomial of degree $(N-M)$ and $B$ is a polynomial \\
of degree~$(M-1)$.
}

\vspace{0.3cm}

\noindent
{\bf Proof.}
We have
$$
( L^{\frac{1}{N-M}+n} )_{\ge 0} = [L(L^{\frac{1}{N-M}+(n-1)})_{\ge 0} +
                        L(L^{\frac{1}{N-M}+(n-1)})_{<0}]_{\ge 0} 
$$
So,
$$
(  L^{\frac{1}{N-M}+n}  )_{\ge 0} = L(L^{\frac{1}{N-M}+(n-1)})_{\ge 0} +
(L( L^{\frac{1}{N-M}+(n-1)} )_{<0})_{\ge 0} - 
$$
$$
(L( L^{\frac{1}{N-M}+(n-1)} )_{\ge 0})_{<0}. 
$$
If we take
\begin{equation}
R_n=(L( L^{\frac{1}{N-M}+(n-1)})_{<0})_{\ge 0} \:
 - (L(L^{\frac{1}{N-M}+(n-1)})_{\ge 0})_{<0} \quad , 
\label{Rn}
\end{equation}

\noindent
then 
$$
( L^{\frac{1}{N-M}+n})_{\ge 0} = 
L( L^{\frac{1}{N-M}+(n-1)} )_{\ge 0}\: + R_n .
$$
Hence,
$$
\frac{\partial L}{\partial t_n}=
\left \{ ( L^{\frac{1}{N-M}+n})_{\ge 0};L \right\}=
\left\{ L(L^{\frac{1}{N-M}+(n-1)} )_{\ge 0} + R_n ; L \right\}
= L\frac{\partial L}{\partial t_n} +\{R_n;L\} , 
$$
and (\ref{SymmRel}) is satisfied.
The remainder $R_n$ has form
(\ref{FormRec}). Indeed  in (\ref{Rn}) we set
$$
A=(L( L^{\frac{1}{N-M}+(n-1)})_{<0})_{\ge 0}
$$
 and
$$
B=\Delta_2\cdot
 (L(L^{\frac{1}{N-M}+(n-1)})_{\ge 0})_{<0}
$$ 
Then $A$ is a polynomial of degree $(N-M-1)$ and $B$ is a polynomial \\
of degree~$(M-1)$.
$\Box$

Now we can apply the Lemma 1 to find recursion operators.

\section*{III. Examples.}

\noindent
{\bf Example 2. } 
Let us consider the equation (\ref{Eqn0}) 
given in introduction.

\vspace{0.3cm}

\noindent
{\bf Lemma 3. } {\it A recursion  operator for  (\ref{Eqn0})
is given by (\ref{Rec1}).}

\vspace{0.3cm}

\noindent
{\bf Proof.}
Using (\ref{FormRec}) for $R_n$,
we have $R_n=A+\displaystyle{\frac{B}{p+Q}}$.
So, the symmetry relation (\ref{SymmRel}) is
$$
\frac{\partial S}{\partial t_n} +
\frac{\partial P}{\partial t_n}\cdot\frac{1}{p+Q} +
\frac{\partial Q}{\partial t_n}\cdot\frac{P}{(p+Q)^2} =
$$
$$
\left ( p+S+\frac{P}{p+Q}\right )
\left ( \frac{\partial S}{\partial t_{n-1}} +
\frac{\partial P}{\partial t_{n-1}}\cdot\frac{1}{p+Q} +
\frac{\partial Q}{\partial t_{n-1}}\cdot\frac{P}{(p+Q)^2} \right ) +  
$$
$$   
p\left ( A_x+\frac{B_x}{p+Q}+\frac{-BQ_x}{(p+Q)^2} \right )
 \left ( 1 + \frac{-P}{(p+Q)^2} \right)
$$
$$  
 - \frac{pB}{(p+Q)^2} 
  \left ( S_x+\frac{P_x}{p+Q}+\frac{-PQ_x}{(p+Q)^2} \right )
$$

\noindent
To have the equality the coefficients of $p$ and $(p+Q)^{-3}$
 must be zero. It gives  the recursion relations to find 
$A$ and $B$.
Then the coefficients of $ p^0,\; (p+Q)^{-1},\; (p+Q)^{-2}$
give  expressions for 
$\displaystyle{\frac{\partial S}{\partial t_n},\;
\frac{\partial P}{\partial t_n},\;
\frac{\partial Q}{\partial t_n} }$.
$\Box$

\vspace{0.3cm}

\noindent{\bf Example 4.}
The Lax equation (\ref{LaxEqn}) with the Lax function (\ref{LF2}), for $n=1$, 
gives the following system

\begin{equation}
\begin{array}{lll}
S_t & = & P_x + Q_x, \\
P_t & = & PS_x,\\
Q_t & = &  QS_x -FQ_x -QF_x, \\
F_t & = & FS_x -FF_x. \\
\end{array}
\label{Eqn2}
\end{equation}

\vspace{0.3cm}

\noindent
{\bf Lemma 5. } {\it A recursion  operator for  (\ref{Eqn2})
is given by 
\begin{equation}
\left(
\begin{array}{llll}
S & 2+P_xD_x^{-1}\cdot P^{-1} & 1 & QF^{-1}+Q_xD_x^{-1}\cdot F^{-1}\\ 
 & & & \\
 2P & S+QF^{-1} +PS_xD_x^{-1}\cdot P^{-1}& PF^{-1} & -2PQF^{-2} \\
  & PF^{-1}(Q_x-QF^{-1}F_x)D_x^{-1}\cdot P^{-1}  & &
-PF^{-1}(Q_x-QF^{-1}F_x)D_x^{-1}\cdot F^{-1}  \\ 
 & & & \\
2Q & -QF^{-1} & S-F & -2PQF^{-2}-2Q \\  
   & -PF^{-1}(Q_x-QF^{-1}F_x)D_x^{-1}\cdot P^{-1} & -PF^{-1}&
                                +PF^{-1}(Q_x-QF^{-1}F_x)D_x^{-1}\cdot F^{-1} \\
 & & & +(QS_x-QF_x-FQ_x)D_x^{-1}\cdot F^{-1} \\   
 & & & \\
F & 1+(P_x-PF^{-1}F_x)D_x^{-1}\cdot P^{-1}  & -1 &
 PF^{-1}-F+(FS_x-FF_x)D_x^{-1}\cdot F^{-1}  \\
  &  &  & -(P_x-PF^{-1}F_x)D_x^{-1}\cdot F^{-1} \\ 
\end{array}
\right ) .
\label{Rec3}
\end{equation} 
}

\vspace{0.3cm}

\noindent
{\bf Proof.}
Using (\ref{FormRec}) for $R_n$,
we have $R_n=C+\displaystyle{\frac{A}{p}+\frac{B}{p+F}}$.
So, the symmetry relation (\ref{SymmRel}) is
$$
\frac{\partial S}{\partial t_n} +
\frac{\partial P}{\partial t_n}\cdot\frac{1}{p} +
\frac{\partial Q}{\partial t_n}\cdot\frac{1}{(p+F)} +
\frac{\partial F}{\partial t_n}\cdot\frac{-Q}{(p+F)^2} =
$$
$$
\left ( p+S+\frac{P}{p}+\frac{Q}{p+F}\right )
\left ( \frac{\partial S}{\partial t_{n-1}} +
\frac{\partial P}{\partial t_{n-1}}\cdot\frac{1}{p} +
\frac{\partial Q}{\partial t_{n-1}}\cdot\frac{1}{(p+F)}  +  
\frac{\partial F}{\partial t_{n-1}}\cdot\frac{-Q}{(p+F)^2} \right )+
$$
$$   
p\left (\frac{-B}{p^2}+\frac{-C}{(p+F)^2} \right )
 \left (S_x + \frac{P_x}{p}+ \frac{Q_x}{(p+F)}+
 \frac{-QF_x}{(p+F)^2} \right)-  
$$
$$ 
 p\left( A_x+\frac{B_x}{p}+\frac{C_x}{(p+F)+\frac{-CF_x}{(p+F)^2}} \right )
  \left (1+\frac{P}{p}+\frac{-Q}{(p+F)^2}  \right )
$$
Therefore, the coefficients of $p$, $p^{-2}$, and $(p+F)^{-3}$
 must be zero, it gives  recursion relations to find 
$A,\; B$ and $C$.
Then the coefficients of $ p^0,\;p^{-1},$ \\
$(p+F)^{-1}$ and $(p+F)^{-2},\;$
give  expressions for 
$\displaystyle{
\frac{\partial S}{\partial t_n},\;
\frac{\partial P}{\partial t_n},}$
${\displaystyle
\frac{\partial Q}{\partial t_n}}$
and $\displaystyle{\frac{\partial F}{\partial t_n}.}$
$\Box$

\section*{Acknowledgments}

I thank Professors Metin G{\" u}rses, Atalay Karasu and Maxim Pavlov 
for several discussions.
This work is partially supported by the Scientific and Technical
Research Council of Turkey.


\begin{thebibliography}{100}

\bibitem{GKS} 
M. G\"urses, A. Karasu,  V.V. Sokolov, {\it J. Math. Phys},
{\bf 40}, 6473-6490 (1999).
\bibitem{B} 
M.Blaszak " On construction
of recursion  operator and algebra of symmetries for field and 
lattice systems", to appear in {\it Rep. Math. Phys}.
\bibitem{GZh} 
M. G{\" u}rses and K. Zhelthukin, {\it J. Math. Phys.}
{\bf 42}, 1309-1325 (2001).
\bibitem{Str} 
I.A.B. Strachan, {\it
J. Math. Phys.}, {\bf 40}, 5058-5079 (1999). 
\bibitem{StrF} 
D.B. Fairlie and I.A.B. Strachan,
 {\it Inverse Problems}, {\bf 12}, 885-908 (1998).
\bibitem{BGZ} 
J.C. Brunelli, M. G{\" u}rses, and K. Zhelthukin,  {\it
Reviews in Mathematical Physics}, vol. 13, No. 4, 529-543 (2001).
\bibitem{BurD}
J.C. Brunelli and A. Das, {\it Phys. Lett. A },235 , 597-602 (1997).
\bibitem{L}
Luen-Chau Li, {\it Commun. Math Phys.}, 203, 573-592 (1999).

\end{thebibliography}
\end{document}